\documentclass[review]{elsarticle}

\usepackage{hyperref}

\usepackage{amssymb}

\newtheorem{theorem}{Theorem}[section]

\journal{Applied Mathematics Letters}

\begin{document}

\begin{frontmatter}

\title{Comments on {\it A note on stability of fractional logistic maps}, Appl. Math. Lett. 125 (2022) 107787}

\author{Mark Edelman\fnref{myfootnote}}
\address{Stern College for Women, Yeshiva University, 245 Lexington Ave., New York, NY 10016}
\fntext[myfootnote]{Email address: edelman@cims.nyu.edu}

\begin{abstract}
In this paper we show that the stability analysis in the paper
{\it A note on stability of fractional logistic maps}, Appl. Math. Lett. 125 (2022) 107787 is incorrect and repeat a proof of a theorem on convergence of a convolution of the product of a converging series with 
a converging sequence. We also mention that the paper commented on should have been edited more carefully.
\end{abstract}


\end{frontmatter}


\section{The main results of the article commented on}

The paper commented on \cite{LS} is an attempt to obtain at least some of the sufficient stbility conditions for at least one fractional map. This is an important problem, which is not solved yet. 

\subsection{Results of the prior investigations}

The main difficulty here is the slow,
as a power law (as $n^{-\alpha}$, where $n$ is the number of iterations and $\alpha$ is the fractional order of a map), convergence of trajectories to stable periodic points, which was analyzed in \cite{MEPL} (Section 1 and Fig. 1), \cite{MECNSNS} (see Fig. 1), \cite{METaieb} (Section 3.2), \cite{MESpringer} (Section 3.3.3 and figures in it), and \cite{MEChaos2014} (see Fig. 5 for $0< \alpha<1$)  for the fractional standard map. For the fractional logistic map with $0< \alpha<1$ the power-law convergence of trajectories is shown in \cite{MEDie} (see Fig. 6.8). The first attempt to derive stability conditions for fractional maps was done in \cite{MEPL} for the fractional standard map. The method of generating functions was proposed in \cite{MEPL} to analyze the stability of fractional maps but the stability conditions for fractional maps have never been obtained.

It is possible to find the equations which define asymptotically periodic points for any fractional maps (see \cite{MEND}). For the asymptotically period two ($T=2$) trajectories in fractional logistic map, the equations were obtained in \cite{MEChaos2018} (see Eq.~(60)--Eq.~(62)). These equations do not have real solutions for the values of the parameter
of the logistic map $\lambda$ (see the equation for the fractional logistic map below)
\begin{eqnarray}
1-\lambda_{C1s}(\alpha)<\lambda<1+ \lambda_{C1s}(\alpha),
\label{FrLogStab}
\end{eqnarray}
where $ \lambda_{C1s}(\alpha)$ is the value of $\lambda$ at which the first bifurcation (from the fixed point to the asymptotically period two point) occurs in the fractional standard map (for details see \cite{MEChaos2018}, see the exact expression for $\lambda_{C1s}$ in Section~\ref{Conc} of this paper). It is confirmed numerically that the fixed point in the logistic map is stable exactly withing the interval of parameters where there are no real asymptotically $T=2$ points, and when the asymptotically period two point appears, this point is stable and the fixed point becomes unstable. But, so far, there is no analytic proof.

For their stability analysis, the authors (Jessica Mendiola-Fuentes and Daniel Melchor-Aguilar) selected the logistic map introduced in \cite{MEAND}
\begin{eqnarray}
X_{n+1}= X_0
+\frac{1}{\Gamma(\alpha)}\sum^{n}_{k=0} \frac{\lambda X_k(1-X_k)-X_k}
{(n+1-k)^{1-\alpha}},
\label{FrCMapx}
\end{eqnarray}
where  $\lambda \in \mathbb{R}$, $0 < \alpha \le 0$,  $X_n=X(t=n)$, $t$ is time, $n \in \mathbb{N}_0$. 
The authors selected a subinterval $1-\Gamma_(\alpha)<\lambda<1+ \Gamma(\alpha)$ of the known interval of stability of the fixed point Eq.~(\ref{FrLogStab}) for their proof of stability. Even if their proof were correct, this wouldn't be sufficient to claim that the bifurcation, fixed point - period two point, can be modeled by the dashed line 
$\lambda=1+\Gamma(\alpha)$ in Fig.~2. The correct bifurcation diagrams, which coincide with the authors' solid line on Fig.~2 can be found in \cite{MEChaos2018} (Fig. 3b) and \cite{MEND} (Fig. 1, and Table 5, where the bifurcation points are given with seven significant figures). There are no real asymptotically period two points between the solid and dashed lines in Fig. 2.

\subsection{The major flaws}

The authors rewrite the linearized equations for the perturbations $x_n=X_n-x_f$ from the fixed points $x_f$ in the form  
\begin{eqnarray}
x_{n+1}= 
\sum^{n-1}_{k=0} [a_0(n-k,\alpha)- a_0(n-k-1,\alpha)]x_k+a_0(0,\alpha)x_n,
\label{Xper}
\end{eqnarray}
where 
\begin{eqnarray}
a_{0}(n,\alpha)=1+\frac{\lambda-1}{\Gamma(\alpha)(n+1)^{1-\alpha}}.
\label{a0}
\end{eqnarray}
The authors show that if $x_0$ is positive, then $x_n$ is positive and bounded and show that  
\begin{eqnarray}
a_{0}(n,\alpha)- a_{0}(n-1,\alpha) \rightarrow 0 \  \ {\rm when} \  \ n \rightarrow \infty.
\label{a0dif}
\end{eqnarray}
Then they claim (without a proof) that for large values of $n$ the total in Eq.~(\ref{Xper}) may be ignored and that in the limit $n \rightarrow \infty $  
\begin{eqnarray}
x_{n+1}=a_0(n,\alpha)x_n.
\label{xlim}
\end{eqnarray}
If true, this would imply the exponential convergence of trajectories to the fixed point and non-zero Lyapunov exponent of the fractional logistic map and contradict to all known numerical results. To show that the assumption made by the authors is incorrect we will employ the following theorem, implicitly proven in \cite{MEND} (we present the explicit proof in the following section):

\begin{theorem}\label{Th1} 
If 
\begin{eqnarray}
\lim_{k \rightarrow \infty}x_k=X,   
\  \ 	 
\lim_{N \rightarrow \infty}\sum_{k=0}^N a_k=S,
\label{The1_1}
\end{eqnarray}
and the series is converging absolutely, then 
\begin{eqnarray}
\Sigma= \lim_{N \rightarrow \infty}\sum_{k=0}^N x_{N-k}a_k=XS,
\label{The1_2}
\end{eqnarray} 
where $k,N \in \mathbb{N}_0$ and $x_k,a_k,X,S,\Sigma \in \mathbb{R}$. 

If $S=\infty$, and $\Sigma$ is finite, then $X=0$.
\end{theorem}
Assuming that the limit $n \rightarrow \infty$ of $x_n$ exists (equal to $X$), the limiting value of the sum in Eq.~(\ref{Xper}) can be written as 
{\setlength\arraycolsep{0.5pt}
\begin{eqnarray}
&&\lim_{n \rightarrow \infty} \sum^{n-1}_{k=0} [a_0(n-k,\alpha) - a_0(n-k-1,\alpha)]x_{n-k}
=\lim_{n \rightarrow \infty}
\sum^{n}_{k=1} [a_0(k,\alpha)
\nonumber    \\
&&- a_0(k-1,\alpha)]x_k =\frac{\lambda-1}{\Gamma(\alpha)}X \lim_{n \rightarrow \infty}\sum^{n}_{k=1}[(k+1)^{\alpha-1}-k^{\alpha-1}]
\nonumber    \\
&&=-X\frac{\lambda-1}{\Gamma(\alpha)}=X[1-a_0(0,\alpha)].
\label{Sum}
\end{eqnarray} 
}
Then, instead of Eq. (\ref{xlim}), in the limit $n \rightarrow \infty $ we will have a trivial equality 
\begin{eqnarray}
x_{n+1}=x_n
\label{xlimn}
\end{eqnarray}
consistent with the power-law convergence.

\section{Proof of the theorem}

A proof of the second statement of Theorem~\ref{Th1} is well described in \cite{MEND} between Eq.~(20) and Eq.~(23). So, here we will prove only the first statement.

Let us consider the difference 
\begin{eqnarray}
&&\Bigl|\Sigma-\sum_{k=0}^N x_{N-k}a_k\Bigr|=
\Bigl| XS-\sum^{N_1}_{k=0} x_{N-k}a_k-
\sum^{N}_{k=N_1+1} x_{N-k}a_k \Bigr|,
\label{The1_2Pr1}
\end{eqnarray} 
where $N>N_1$. If for $\forall<\varepsilon$ we may find $N_l$, such that this difference is less than $\varepsilon$, $\forall N>N_l$, then the theorem is proven.


According to our assumption, $x_n$ is converging and it must be bounded: $|X-x_n|<C1$. Absolute convergence of the total $\Sigma_{k=0}^{\infty} |a_k|$ implies that for any $\varepsilon>0$ there exists $N_{1}$ such that $\Sigma_{k=N_1+1}^{N} |a_k|<\varepsilon/(3C1)$, $\forall N>N_1$.

Because $x_n$ is converging, for any $\varepsilon>0$ we may find $N_{l1}$ such that $|X-x_{N-k}|<\varepsilon/ (3\sum _{k=0}^\infty |a_k|)$, $\forall N>N_{l1}$ and $k \le N_1$.
We may also find $N_{l2}$ such that $\sum _{k=N+1}^\infty |a_k| <\varepsilon/(3*|X|)$. Now, for $N>\max\{N_{l1},N_{l2}\}$, the following chain of inequalities will hold
{ \setlength\arraycolsep{0.5pt}
\begin{eqnarray}
&&\Bigl| XS-\sum^{N_1}_{k=0} x_{N-k}a_k-
\sum^{N}_{k=N_1+1} x_{N-k}a_k \Bigr| 
=\Bigl|\sum_{k=0}^{N_1} 
(X-x_{N-k})a_k
\nonumber    \\
&&-\sum_{k=N_1+1}^{N} (X-x_{N-k})a_k+X\sum_{k=N+1}^{\infty}a_k \Bigr|
< \Bigl|\sum_{k=0}^{N_1} (X-x_{N-k})a_k \Bigr|
\nonumber    \\
&&+ \Bigl|\sum_{k=N_1+1}^{N} (X-x_{N-k})a_k \Bigr|+|X| \Bigl|\sum_{k=N+1}^{\infty}a_k \Bigr|
\\
&&<\frac{\varepsilon}{3\sum _{k=0}^\infty |a_k|}\sum_{k=0}^{N_1} |a_k|+C1\frac{\varepsilon}{3C1}+|X|\frac{\varepsilon}{3|X|} \le \varepsilon.
\label{Ineq}
\end{eqnarray} 
}
This completes the proof of the theorem.

\section{Conclusion}
\label{Conc}

From the investigation of the period-two points (see Eq.~(62) in 
\cite{MEChaos2018} and Eq.~(49) in \cite{MEND}) and numerical simulations,
the $x=0$ fixed point is stable when
\begin{eqnarray}
&&1-\frac{\Gamma(\alpha)}{\sum^{\infty}_{k=0}\Bigl[(2k+1)^{\alpha-1}-
(2k+2)^{\alpha-1}\Bigr]} < \lambda <1 
\label{Stab1}
\end{eqnarray} 
and the $x=(\lambda-1)/\lambda$ fixed point is stable when
\begin{eqnarray}
&&1<\lambda<1+\frac{\Gamma(\alpha)}{\sum^{\infty}_{k=0}\Bigl[(2k+1)^{\alpha-1}-
(2k+2)^{\alpha-1}\Bigr]},
\label{Stab2}
\end{eqnarray} 
where the total in the denominator, called $S_2$ in \cite{MEND}, has values between approximately 0.7, for $\alpha=0$, and 0.5,
for $\alpha=1$ (see Table 3 in \cite{MEND}). The intervals of stability are much wider than those considered in the paper commented on. 

The methods that were previously used to analyze the stability of fractional systems include applications of generating functions \cite{MEPL} and $\mathbb{Z}$-transform (which works better in the case of fractional difference equations, see, e.g., \cite{z,Cermak,Moz}). The generating function $\tilde{X}(t)$ is defined as 
\begin{eqnarray}
&&\tilde{X}(t)= \sum^{\infty}_{n=0}x_nt^n.
\label{Xtilde}
\end{eqnarray} 
Multiplication of Eq.~(\ref{Xper}) by $t^n$ and summation from zero to infinity gives
\begin{eqnarray}
&&\tilde{X}(t)= \frac{x_0}{1-\Bigl[\frac{\lambda-1}{\Gamma(\alpha)}\tilde{J}(t)+ \Bigl(1+ \frac{\lambda-1}{\Gamma(\alpha)} \Bigr) \Bigr]t},
\label{Xtilde1}
\end{eqnarray} 
where
\begin{eqnarray}
&&\tilde{J}(t)= \sum^{\infty}_{j=1}\Bigl[(j+1)^{\alpha-1}-j^{\alpha-1}\Bigr]t^j.
\label{Jtilde}
\end{eqnarray} 
When $\lambda=1$, the coefficients of Taylor series of $\tilde{X}(t)$ are $x_n=x_0$ and it seems that this is a critical value at which stability conditions change but there is no proof. The $\mathbb{Z}$-transform is defined as
\begin{eqnarray}
&&\bar{X}(z)= \sum^{\infty}_{n=0}x_nt^{-n}.
\label{z1}
\end{eqnarray} 
From Eq.~(\ref{Xper})
\begin{eqnarray}
&&\bar{X}(z)= \frac{x_0}{1-\Bigl[\frac{\lambda-1}{\Gamma(\alpha)}\bar{J}(z)+ \Bigl(1+ \frac{\lambda-1}{\Gamma(\alpha)} \Bigr) \Bigr]z^{-1}},
\label{z2}
\end{eqnarray} 
where 
\begin{eqnarray}
&&\bar{J}(z)= \sum^{\infty}_{j=1}\Bigl[(j+1)^{\alpha-1}-j^{\alpha-1}\Bigr]z^{-j}.
\label{z3}
\end{eqnarray} 
The stability conditions Eq.~(\ref{Stab1}) and the power-law convergence are not obtained yet neither from the Taylor expansion of the generating function Eq.~(\ref{Xtilde1}) nor from the application of the inverse $\mathbb{Z}$-transform to $\bar{X}(z)$ Eq.~(\ref{z2}).

The authors brought up an important problem. They made a mistake but, as I know (from the private communication with the authors), they found a different way to investigate stability of the fixed points within the interval $1-\Gamma(\alpha)<\lambda<1+\Gamma(\alpha)$. The original paper should not have been edited more carefully to avoid mistakes.

\section*{Acknowledgements}
The author acknowledges support from Yeshiva University's 2021--2022 Faculty Research Fund and expresses his gratitude to Virginia Donnelly for technical help.

\end{document}